\documentclass[%
 aip,
 amsmath,amssymb,
 reprint,%
]{revtex4-1}

\usepackage{graphicx}
\usepackage{dcolumn}
\usepackage{bm}
\usepackage{svg}

\usepackage[utf8]{inputenc}
\usepackage[T1]{fontenc}
\usepackage{mathptmx}
\usepackage{etoolbox}
\usepackage{comment}

\makeatletter
\def\@email#1#2{%
 \endgroup
 \patchcmd{\titleblock@produce}
  {\frontmatter@RRAPformat}
  {\frontmatter@RRAPformat{\produce@RRAP{*#1\href{mailto:#2}{#2}}}\frontmatter@RRAPformat}
  {}{}
}%
\makeatother
\begin{document}

\preprint{AIP/123-QED}

\title[Self-cooling, blue-detuned dissipative Kerr microresonator soliton comb]{Self-cooling, blue-detuned dissipative Kerr microresonator soliton comb}
\author{Kenji Nishimoto}
 \affiliation{Graduate School of Sciences and Technology for Innovation, Tokushima University, 2-1, Minami-Josanjima, Tokushima, Tokushima 770-8506, Japan
 }
 \affiliation{Institute of Post-LED Photonics, Tokushima University, 2-1, Minami-Josanjima, Tokushima, Tokushima 770-8506, Japan
 }

\author{Kaoru Minoshima}
\affiliation{Institute of Post-LED Photonics, Tokushima University, 2-1, Minami-Josanjima, Tokushima, Tokushima 770-8506, Japan
}%
\affiliation{Graduate School of Informatics and Engineering, The University of Electro-Communications, 1-5-1 Chofugaoka, Chofu, Tokyo 182-8585, Japan
}%

\author{Naoya Kuse}
 \email{kuse.naoya@tokushima-u.ac.jp}
\affiliation{Institute of Post-LED Photonics, Tokushima University, 2-1, Minami-Josanjima, Tokushima, Tokushima 770-8506, Japan
}%
\affiliation{Institute of Photonics and Human Health Frontier, Tokushima University, 2-24, Shinkura-cho, Tokushima 770-8501, Japan
}%

\date{\today}

\begin{abstract}
Dissipative Kerr solitons (DKSs) generated in high-Q microresonators driven by continuous-wave (CW) lasers provide chip-scale optical frequency combs composed of mutually coherent CW lines. However, their small mode volume makes them highly susceptible to thermal fluctuations, and the resulting thermo-refractive noise (TRN) perturbs the repetition rate $f_{\rm rep}$. Here, we experimentally demonstrate a “blue-detuned” DKS in a coupled-ring microresonator. By employing avoided-mode-crossing (AMX)-induced dispersion engineering at the pump mode, DKSs are generated even when the pump laser is tuned to the higher-frequency (blue) side of the resonance. In this regime, the pump laser not only seeds DKS formation but also serves as a cooling laser for the thermally sensitive pumped mode. We observe a self-cooling effect that reduces the phase noise of $f_{\rm rep}$ by up to 14.5 dB, while achieving a pump-to-comb conversion efficiency as high as 37\%. These results establish blue-detuned DKSs as a thermally robust and power-efficient solution for integrated microcomb systems, eliminating the need for auxiliary lasers.
\end{abstract}

\maketitle

\section{Introduction}

\begin{figure*}[htbp!]
\centering
\includegraphics[width=0.9\linewidth]{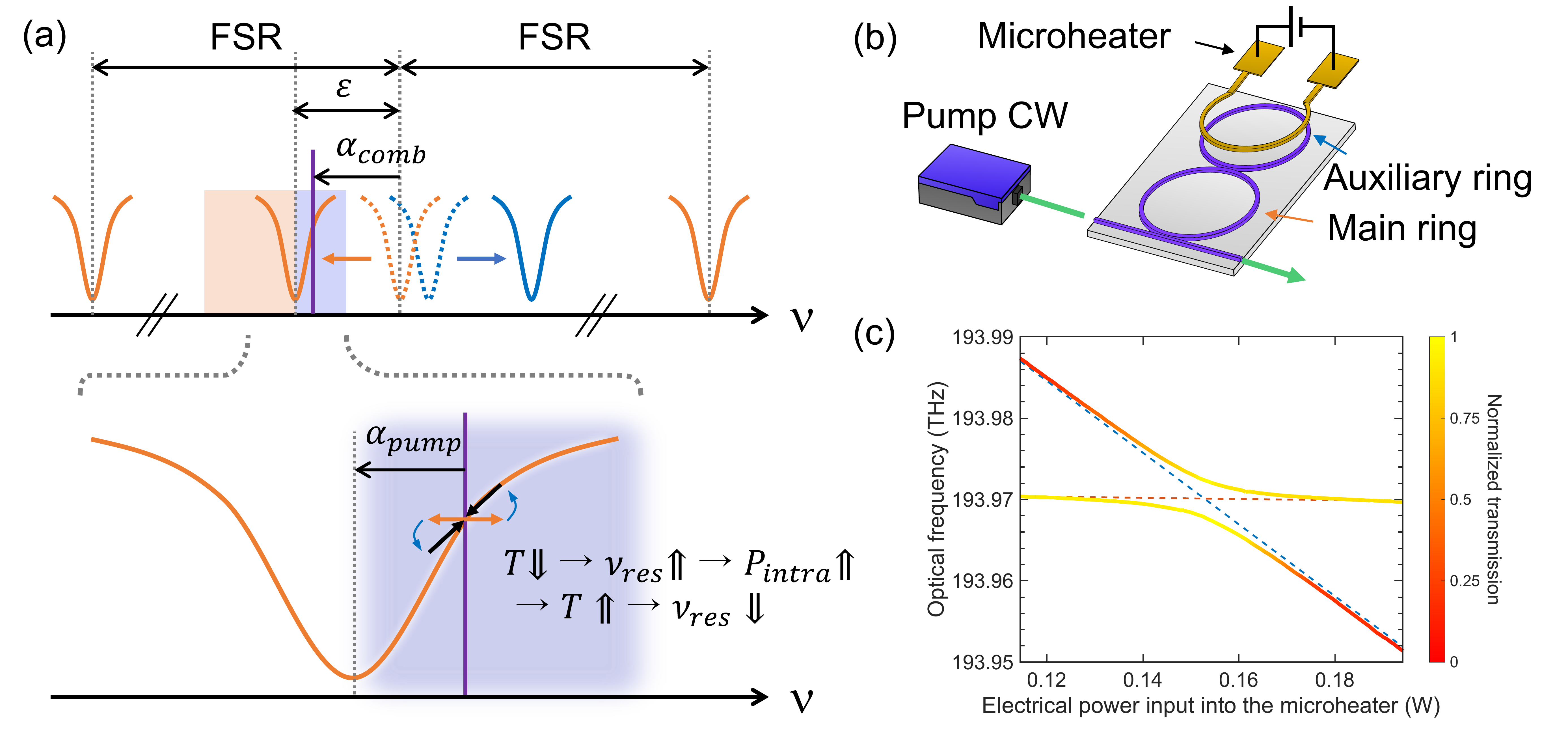}
\caption{(a) Concept of the pump-AMX configuration. Red and blue resonances show the resonances of the main and auxiliary rings, respectively. The purple line shows the frequency of the pump CW laser. At the pump mode, AMX is introduced, and the resonances are repulsed by the amount of $\varepsilon$, modifying the detuning relation as $\alpha_{\rm comb} = \alpha_{\rm pump} + \varepsilon$, which allows $\alpha_{\rm comb} > \sqrt{3}$ while $\alpha_{\rm pump} < 0$, enabling a blue-detuned DKS. For the blue-detuned DKS, the self-cooling effect is expected as shown in the bottom figure. (b) Schematic of the coupled-ring microresonator. The main ring generates the DKS, and the auxiliary ring controls AMX via a microheater. (c) Measured resonance-frequency shifts near $193.97~\mathrm{THz}$ versus heater power. Colored curves show the coupled resonances, dotted lines indicate the uncoupled resonances calculated using Eq.~(\ref{eq:AMX}). }
\end{figure*}

Dissipative Kerr solitons (DKSs), generated by pumping high-Q microresonators with continuous-wave (CW) lasers, have emerged as a chip-scale platform for broadband, phase-stable optical frequency combs \cite{Herr2014, kippenberg2018dissipative}. In recent years, wafer-scale integrated microresonators based on Si$_3$N$_4$ \cite{xiang2022silicon, ji2024efficient} have enabled full on-chip integration of DKSs \cite{xiang2021laser}. Moreover, the extremely short cavity length of microresonators allows mode spacing ($f_{\rm rep}$) in the GHz--THz range \cite{liu2020photonic, helgason2021dissipative, pfeiffer2017octave}, supporting applications such as optical atomic clocks \cite{newman2019architecture, wu2025vernier}, mm/THz wave generation \cite{kudelin2024photonic, kuse2022low, tetsumoto2021optically}, parallelized LiDAR \cite{riemensberger2020massively}, optical communications \cite{marinPalomo2017microresonator, corcoran2025optical}, and optical computing \cite{feldmann2021parallel, Xu2021}.

For mm/THz wireless communication and radar systems that employ mm/THz carriers generated from DKSs, the phase noise of $f_{\rm rep}$ is critical \cite{tokizane2023terahertz, jia2025low, kittlaus2021low}, since the phase noise of mm/THz carriers directly corresponds to that of $f_{\rm rep}$. The phase noise of $f_{\rm rep}$ also limits that of individual comb modes \cite{nishimoto2020investigation, lei2022optical}, which is important for optical communications \cite{corcoran2025optical} and LiDAR \cite{riemensberger2020massively} due to noise magnification relative to the pump mode. This phase noise is fundamentally limited by thermo-refractive noise (TRN) in microresonators \cite{stone2020harnessing, huang2019thermorefractive, nishimoto2020investigation, lei2022optical, yang2021dispersive}, owing to their small mode volume. To suppress the phase noise, both active and passive approaches have been reported. In active schemes, $f_{\rm rep}$ is stabilized to low-noise CW lasers \cite{tetsumoto2021optically} or long fiber delay lines \cite{kuse2022low}. However, such active methods introduce additional components and complexity, undermining the chip-scale advantage of DKS systems. Alternatively, passive approaches have been explored, in which TRN is optically suppressed—so-called self-cooling. In the self-cooling method, an optical carrier is coupled to the blue-detuned side of a resonance \cite{drake2020thermal, nishimoto2022thermal}. The optical carrier can be provided by another laser \cite{drake2020thermal} or by a sideband derived from the pump CW laser \cite{nishimoto2022thermal}. Nevertheless, systems that employ an additional optical carrier suffer from increased complexity and issues such as noise transfer from the cooling laser, unwanted parametric oscillations, and modulation instability \cite{nishimoto2022thermal}. To address these limitations, alternative schemes have been proposed in which the pump laser used for DKS generation is simultaneously coupled to an additional cavity mode to induce cooling \cite{lei2022thermal}. For instance, by exciting modes from different mode families (e.g., TE and TM) within the same resonator, one can generate a DKS in one mode while suppressing thermal fluctuations through another. However, this approach still imposes new technical constraints, including complex polarization control and low yield due to poor tolerance to fabrication errors.

Here, we focus on a localized dispersion shift enabled by photonic-crystal rings (PhC rings) \cite{yu2021spontaneous, yu2022continuum, lucas2023tailoring, moille2023fourier} and coupled-ring microresonators \cite{helgason2021dissipative, tikan2021emergent, helgason2023surpassing}. Such a shift arises from the repulsive interaction between two closely spaced resonances in different mode families, an effect known as avoided mode crossings (AMX) \cite{liu2014investigation, Miller2015}. The resulting local modification of dispersion alters the phase-matching condition and enables deterministic and spontaneous DKS generation \cite{yu2021spontaneous}, substantial enhancement of conversion efficiency \cite{helgason2023surpassing}, spectral shaping andbroadening \cite{moille2018phased, moille2023fourier, nishimoto2023spectral,lucas2023tailoring}, active tuning of dispersive waves \cite{okawachi2022active}, and dark-soliton generation \cite{helgason2021dissipative, yu2022continuum}. When a localized dispersion shift is introduced at the resonance pumped by the CW laser in an anomalous-dispersion microresonator (hereafter referred to as the pump-AMX configuration), the formation mechanism and properties of DKSs are modified because the comb detuning ($\alpha_{\rm comb}$) differs from the pump detuning ($\alpha_{\rm pump}$). Here, $\alpha_{\rm comb}$ is the detuning between the pump CW laser and the original resonance frequency ignoring AMX, and $\alpha_{\rm pump}$ is the detuning between the pump CW laser and the actual resonance with AMX. Both are normalized by half the resonance linewidth. Spontaneous DKS generation that bypasses a chaotic-comb regime is possible when $\alpha_{\rm comb} > 0$ and $\alpha_{\rm pump} < 0$, in which no phase-matching condition for a Turing pattern is satisfied \cite{yu2021spontaneous}. In addition, for $\alpha_{\rm comb} > 0$ and $\alpha_{\rm pump} \approx 0$, the coupling efficiency between the DKS and the pump CW laser is improved, leading to pump-to-comb energy conversion efficiencies approaching $60\%$, well above conventional DKS values (typically a few percent) \cite{helgason2023surpassing}. Furthermore, the optical spectrum of a DKS can be broader than that of a conventional DKS because $\alpha_{\rm comb}$ can be larger than $\alpha_{\rm pump}$ ($> 0$) \cite{nishimoto2023spectral}. Throughout this paper, we refer to a DKS with $\alpha_{\rm pump} > 0$ as a "red-detuned" DKS and one with $\alpha_{\rm pump} < 0$ as a "blue-detuned" DKS.

In this study, we report self-cooling that requires neither an additional optical carrier nor complex resonance-mode management, by harnessing a blue-detuned DKS in a coupled-ring microresonator. In this configuration, the pump CW laser serves not only to generate the DKS but also to provide cooling. Localized and selective dispersion control via AMX in the coupled-ring microresonator supports deterministic pulse formation of a blue-detuned DKS. Experimentally, we confirm that the blue-detuned DKS exhibits more than $10~\mathrm{dB}$ lower phase noise in $f_{\rm rep}$ than a conventional red-detuned DKS, while achieving a much higher ($\approx 37\%$) pump-to-comb conversion efficiency. In addition, we thoroughly investigate the methods and dynamics required to access blue-detuned DKSs in coupled-ring microresonators, which differ markedly from those in PhC rings.

\section{Concept and experimental setup}
In our approach, a blue-detuned DKS is generated from a self-cooled microresonator. The pump-AMX configuration is realized by employing a PhC ring or a coupled-ring microresonator. The key idea is to generate a DKS with $\alpha_{\rm pump} < 0$. For standard microresonators without the pump-AMX configuration, $\alpha_{\rm pump}$ is always equal to $\alpha_{\rm comb}$, and $\alpha_{\rm pump}$ must exceed $\sqrt{3}$ to support bistability between a flat background field and a soliton pulse \cite{godey2014stability}. Thus, only a red-detuned DKS can be generated. In contrast, in the pump-AMX configuration, $\alpha_{\rm comb}$ is modified as $\alpha_{\rm comb} = \alpha_{\rm pump} + \varepsilon$, as shown in Fig.~1(a). Here, $\varepsilon$ is the normalized resonance shift, defined as the frequency deviation of the coupled resonances from the original resonances divided by half the resonance linewidth. This allows the condition $\alpha_{\rm comb} > \sqrt{3}$ to be satisfied for the DKS while $\alpha_{\rm pump} < 0$ \cite{yu2021spontaneous, helgason2023surpassing}. In this case, the pump CW laser is located at the high-frequency side (blue side) of the lower-frequency resonance of the resonance pair split by AMX. Then, similar to previous self-cooling demonstrations \cite{drake2020thermal, nishimoto2022thermal, lei2022thermal}, the microresonator is cooled as follows. A resonance frequency fluctuation $\delta\nu_{\rm res}$ caused by a thermal fluctuation $\delta T$ leads to a simultaneous fluctuation in the intracavity power $\delta P_{\rm intra}$. The thermal shift induced by $\delta P_{\rm intra}$ acts in the opposite direction to $\delta\nu_{\rm res}$, generating a photothermal force that passively suppresses the thermal fluctuation of the resonance. As a result, a blue-detuned DKS can naturally exhibit a self-cooling effect under appropriate conditions, and this effect can coexist with the high conversion efficiency reported in Ref.~\cite{helgason2023surpassing}.

In our system, AMX is induced by employing a coupled-ring microresonator. A schematic illustration of the coupled-ring microresonator used in the experiment is shown in Fig.~1(b). The device consists of Si$_3$N$_4$ waveguides. The main ring for DKS generation and the auxiliary ring for AMX control are linearly coupled. Their free spectral ranges (FSRs) are 950~GHz and 995~GHz, and their loaded quality factors are $7.8\times10^5$ and $8.4\times10^5$, respectively. The FSR difference of 45~GHz between the two rings is used to induce only one AMX within the optical spectrum of a DKS. A microheater is placed on the auxiliary ring. When the input power to the heater is tuned, only the resonance of the auxiliary ring exhibits a large frequency shift due to the photothermal effect. Figure~1(c) shows the resonance frequency shifts of both the main and auxiliary rings near 193.97~THz as a function of the heater power, which are measured using a wavelength meter while monitoring the transmitted optical power. As the heater power increases, the two resonances (colored curves) approach each other and deviate from their uncoupled positions (dotted lines), indicating the presence of an AMX characterized by mutual repulsion. The dotted lines are calculated as
\begin{gather}
f_{m,a} = \frac{f_{+} + f_{-}}{2} \pm \sqrt{\frac{(f_{+} - f_{-})^{2}}{4} - \kappa_{m-a}^{2}}
\label{eq:AMX}
\end{gather}
Here, $f_{m,a}$ denote the uncoupled resonance frequencies of the main and auxiliary mode families, respectively, while $f_{+}$ and $f_{-}$ represent the measured higher- and lower-frequency resonances under mode coupling. $\kappa_{m-a}$ is the mode-splitting strength, defined as the frequency difference between $f_{+}$ and $f_{-}$ at their closest separation. From Fig.~1(c), the mode-splitting strength is estimated to be $\kappa_{m-a} \approx 5.2$~GHz, resulting in a maximum value of $\varepsilon$ of about 20.

\begin{figure}[t]
\centering
\includegraphics[width=0.9\linewidth]{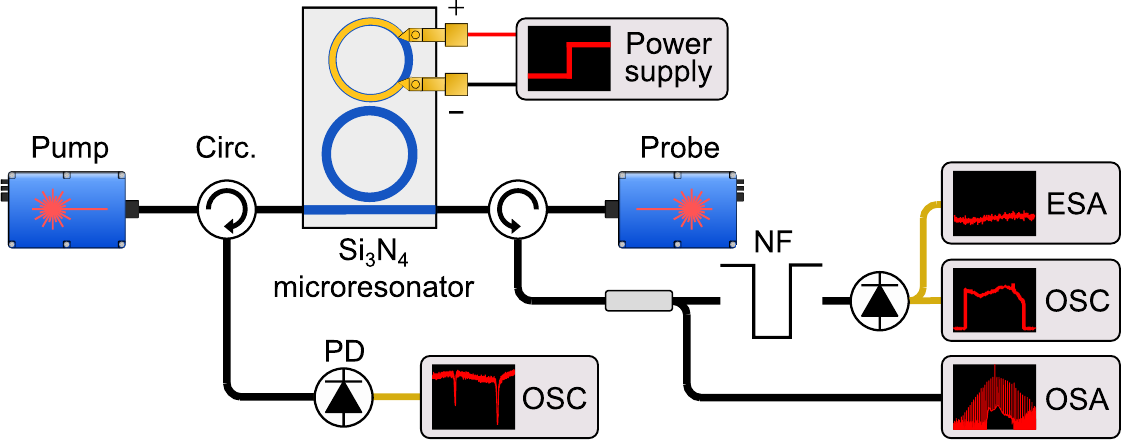}
\caption{Schematic of the experimental setup. Two independent CW lasers were used for the pump and probe. Circ.: optical circulator, PD: photodetector, OSC: oscilloscope, NF: optical notch filter, ESA: electrical spectrum analyzer, OSA: optical spectrum analyzer.}
\end{figure}

Figure~2 shows the experimental setup. A pump CW laser with an emission wavelength of $1546~\mathrm{nm}$ is injected into the coupled-ring microresonator to generate a blue-detuned DKS. The input optical power is between $10$ and $100~\mathrm{mW}$. The output from the coupled-ring microresonator is split and used to measure optical and RF spectra and to monitor the formation of the blue-detuned DKS. For the latter two measurements, the pump CW laser is suppressed with a notch filter (NF). In addition, an independent probe CW laser is injected from the counter-propagating direction. The probe CW laser power is sufficiently low so that it does not affect the generation of the blue-detuned DKS. The probe CW laser is used to measure the resonance transmission spectrum of the coupled-ring microresonator influenced by AMX and to record the beat signal with the pump CW laser. These measurements allow us to estimate local resonance frequency shifts and effective detuning.

\begin{figure*}[t]
\centering\includegraphics[width=1.0\linewidth]{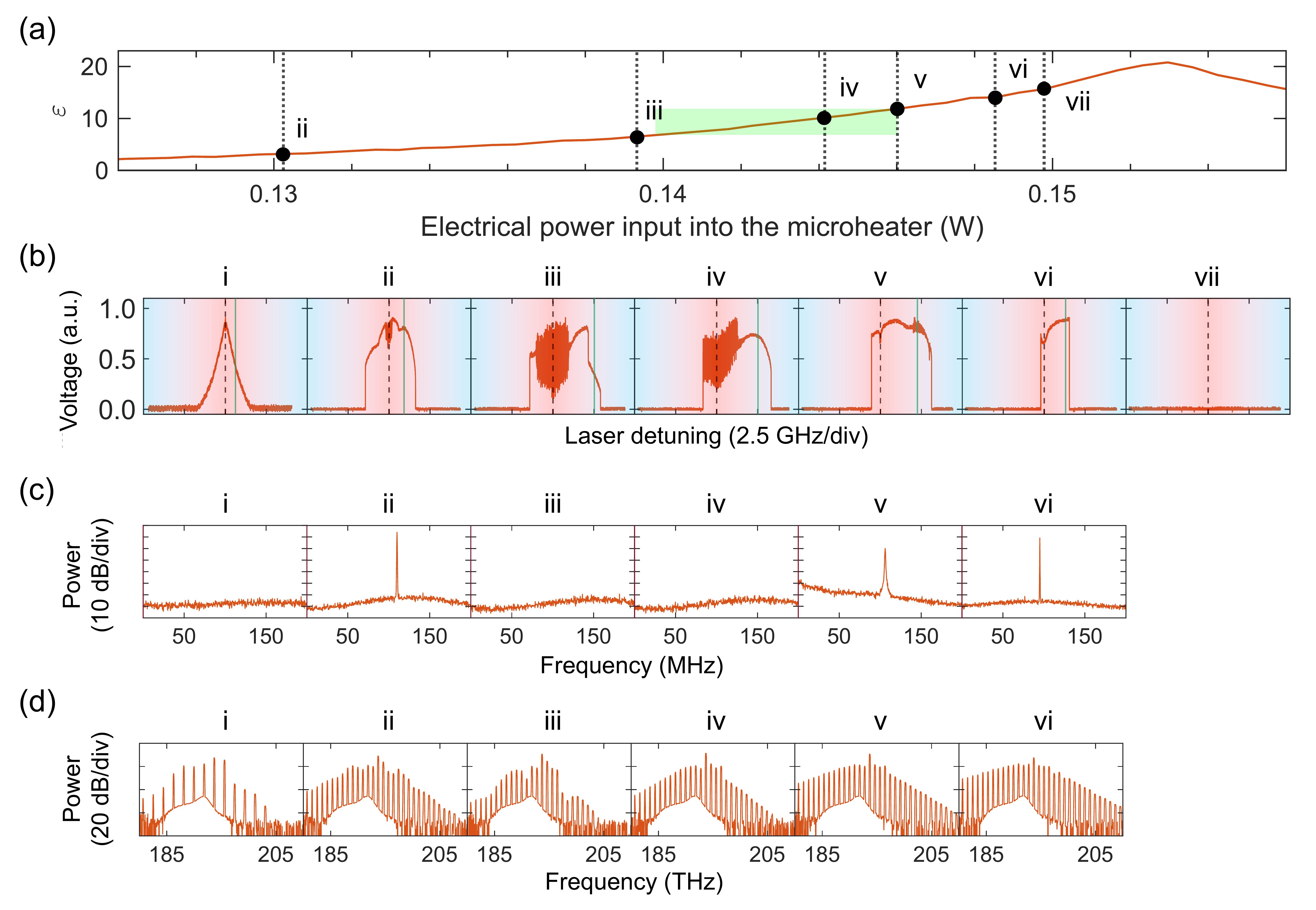}
\caption{Dependence of blue-detuned DKS generation on $\varepsilon$. (a) Measured $\varepsilon$ as a function of heater power. Black circles labeled ii to vii mark the $\varepsilon$ values where comb-power evolution, RF and optical spectra are analyzed.  The green highlighted region indicates the range where blue-detuned DKSs without breathing oscillations are accessed.(b) Comb power excluding the pump versus detuning. Dotted lines indicate where the scan direction is reversed. The pump frequency is scanned from blue to red, then from red to blue. State i corresponds to $\varepsilon = 0$. (c) RF noise spectra and (d) optical spectra at the detunings marked by green lines in (b) for states i to vi.}
\end{figure*}

\section{Generation of blue-detuned DKS}
The generation of a blue-detuned DKS is investigated by varying $\varepsilon$ and scanning the pump-laser frequency to determine the appropriate $\varepsilon$ value. The power of the pump CW laser $P_{\rm pump}$ is fixed at $50~\mathrm{mW}$. Figure~3(a) shows the measured change in $\varepsilon$ as a function of the heater power applied to the microheater on the auxiliary ring. The black lines labeled ii through vii in Fig.~3(a) indicate the $\varepsilon$ values at which the change in comb power, excluding the pump CW laser, is investigated, as shown in Fig.~3(b). The dotted lines in Fig.~3(b) indicate the points at which the direction of the frequency scan is reversed. First, the frequency of the pump CW laser is scanned from blue to red. Then, the scan direction is reversed. The turning point is defined as the detuning just before the comb vanishes when the pump CW laser enters the red side of the resonance. Consequently, we find that a DKS is generated only when the frequency is scanned from red to blue after the initial blue-to-red scan. In Fig.~3(b), the state when $\varepsilon = 0$ is labeled as i. Figures~3(c) and 3(d) show the RF noise spectra and optical spectra of the combs obtained at the detuning marked by the green lines in Fig.~3(b) for each state i to vii. At state i ($\varepsilon = 0$), cascaded four-wave mixing occurs, and even after transitioning to a Turing-pattern state \cite{qi2019dissipative, godey2014stability}, no further transition to another intracavity field state is observed. As $\varepsilon$ increases ($\varepsilon \approx 3.0$), a breathing state is spontaneously reached without passing through Turing or chaotic combs, as highlighted in Fig.~3(b)-ii. In Fig.~3(d)-ii, comb modes spaced by the FSR are observed, and oscillations appear in the RF spectrum, as shown in Fig.~3(c)-ii. As $\varepsilon$ is further increased ($\varepsilon \approx 5.8$), a quiet comb without oscillations is observed, as highlighted in Fig.~3(b)-iii. However, the optical spectrum of this quiet comb does not exhibit a smooth $\mathrm{sech}^2$ envelope. When $\varepsilon$ is gradually increased from the value shown in Fig. 3(b)-iii to $\approx$ 9.5, the comb-mode power of the quiet comb increases, and a blue-detuned DKS is finally generated, as shown in Fig.~3(b)-iv. The blue-detuned DKS can be generated for $\varepsilon$ values within the green highlighted area in Fig.~3(a). The corresponding state transitions are examined in detail in Fig.~4 and discussed in the next section. When $\varepsilon$ is further increased ($\varepsilon > 11.2$), no blue-detuned DKS is obtained, and only the breathing region remains [Fig. 3(b)-v and vi]. When $\varepsilon$ exceeds approximately $15$ [Fig.~3(a)-vii], no comb is observed because $\alpha_{\rm comb}$ becomes too large. From iii to vi, once a breathing state is spontaneously generated by blue-to-red scanning, the detuning range on the blue side that maintains the comb is extended, which is probably due to a bifurcation between the CW and comb states \cite{Herr2014}. These results confirm that by appropriately adjusting $\varepsilon$, which ranges from $6.2$ to $11.2$ at $P_{\rm pump} = 50~\mathrm{mW}$ in our device, a stable single blue-detuned DKS is generated without breathing oscillations even under a static, slow frequency scan. Furthermore, generation of the blue-detuned DKS under these conditions is highly reproducible, as experimentally verified through 120 consecutive accesses (see Supplementary Section~1).

\begin{figure*}[t]
\centering\includegraphics[width=0.9\linewidth]{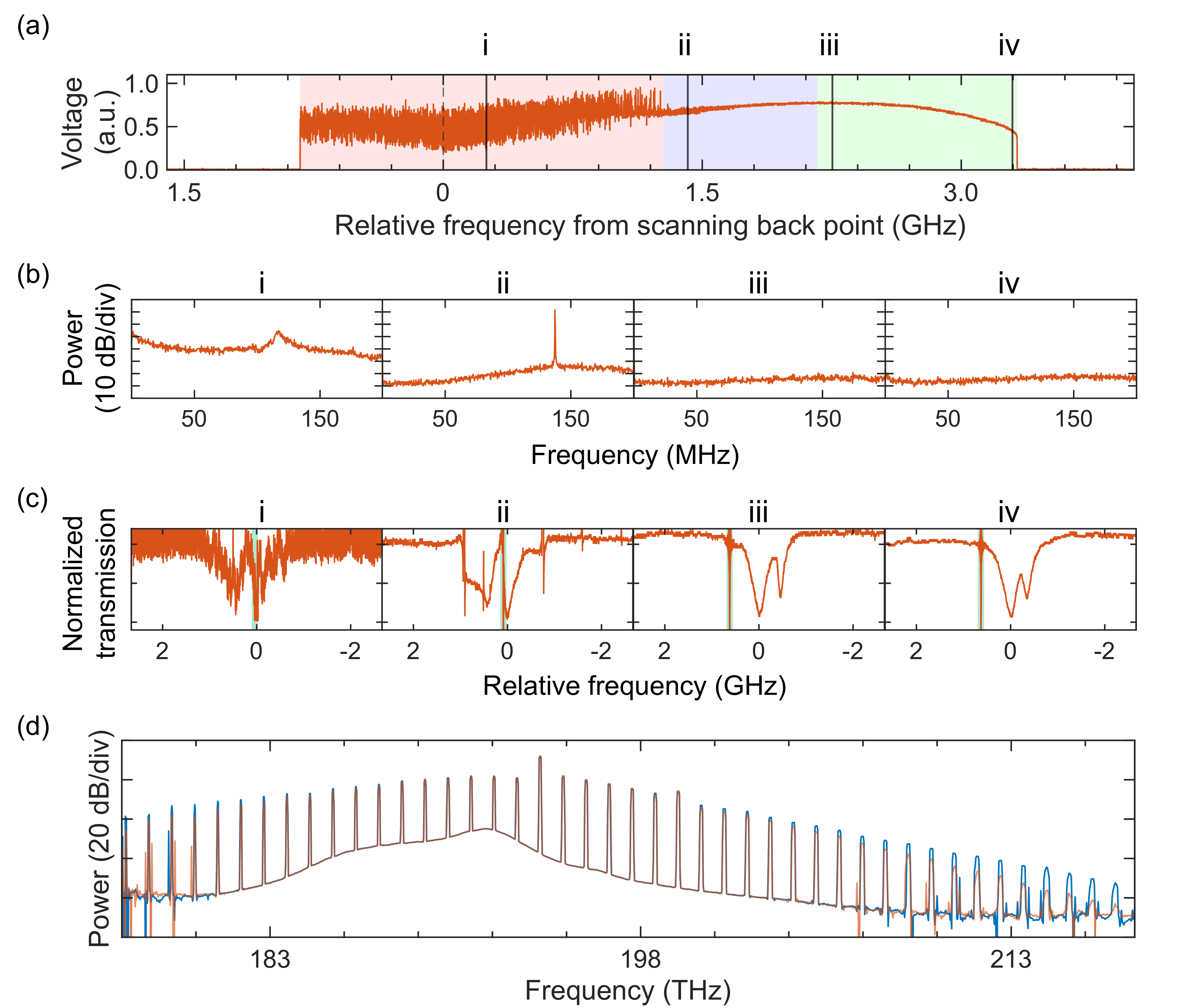}
\caption{(a) Comb power variation with $\varepsilon \approx 9.5$. At the dashed line, the direction of the frequency scan is reversed from blue-to-red to red-to-blue. The regions highlighted in red, blue, and green indicate the non-periodic breathing state, periodic breathing state, and blue-detuned DKS, respectively. Lines i through iv mark the detunings at which the RF spectra and transmission are measured, as shown in (b) and (c). (b) RF noise spectra of the combs measured at points i through iv in (a). (c) Transmission spectra around the pumped resonance, measured at points i through iv in (a). The green squares highlight the beat signals between the pump CW laser and the probe laser. (d) Optical spectra of the blue-detuned DKS (red) and the non-periodic breathing state (blue).}
\end{figure*}

\section{Characterization of blue-detuned DKS}
Figure~4 shows a detailed investigation of the intracavity-field transitions under the conditions observed in Fig.~3(b)-iv ($P_{\rm pump} = 50~\mathrm{mW}$, $\varepsilon \approx 9.5$). Figure~4(a) shows the evolution of the comb power at different detunings. The RF noise spectra and transmission spectra measured with a probe CW laser at points i to iv in Fig.~4(a) are shown in Figs.~4(b) and 4(c), respectively. In Fig.~4(a), an abrupt increase in comb-mode power (around $0.83~\mathrm{GHz}$) is observed during the frequency scan from blue to red. This corresponds to a state dominated by breathing oscillations with low periodic purity (red area in Fig.~4(a)). The RF spectrum of this state is broad, as shown in Fig.~4(b)-i. Here, we distinguish this state from a chaotic state, because the chaotic-like breathing state observed here continuously transitions to a standard breathing state (state ii in Fig.~4(a)). In addition, the optical spectrum shows a smooth $\mathrm{sech}^2$ envelope (Fig.~4(d), blue curve). These observations suggest that state i corresponds to breathing oscillations occurring near the lower-detuning boundary of the breathing state \cite{helgason2023surpassing}. Reversing the frequency scan direction from red to blue leads to a breathing state with narrower RF peaks (around 1.28 GHz; this region is highlighted in blue in Fig. 4(a)). Further scanning toward higher frequency causes the breathing oscillations to vanish, and the system moves into a blue-detuned DKS (green area in Fig.~4(a)). The blue-detuned DKS has a smooth $\mathrm{sech}^2$ envelope in the optical spectrum (Fig.~4(d), red curve), but its spectral bandwidth is narrower than that of state i. This reduction indicates a smaller $\alpha_{\rm comb}$ for the blue-detuned DKS. As evidence for blue-detuned operation, the positions of the pump CW laser and the pumped resonance are shown in Fig.~4(c). In Fig.~4(c), two resonances are observed: the pumped resonance on the blue side and the soliton resonance on the red side \cite{guo2017universal}. When the blue-detuned DKS is generated, the pump CW laser is clearly detuned to the blue side of the pumped resonance. The blue-detuned DKS remains stable until the detuning reaches the blue-side edge of the soliton existence range (iv in Fig.~4(a)). All states shown in Fig.~4(a) can be continuously and reversibly accessed by detuning control. In state iii, the power-conversion efficiency from the pump CW laser to the comb modes, defined as the ratio of the total power of the comb modes (excluding the pump) to the CW laser power injected into the chip \cite{helgason2023surpassing}, is measured to be approximately $37\%$.

\begin{figure*}[t]
\centering\includegraphics[width=1.0\linewidth]{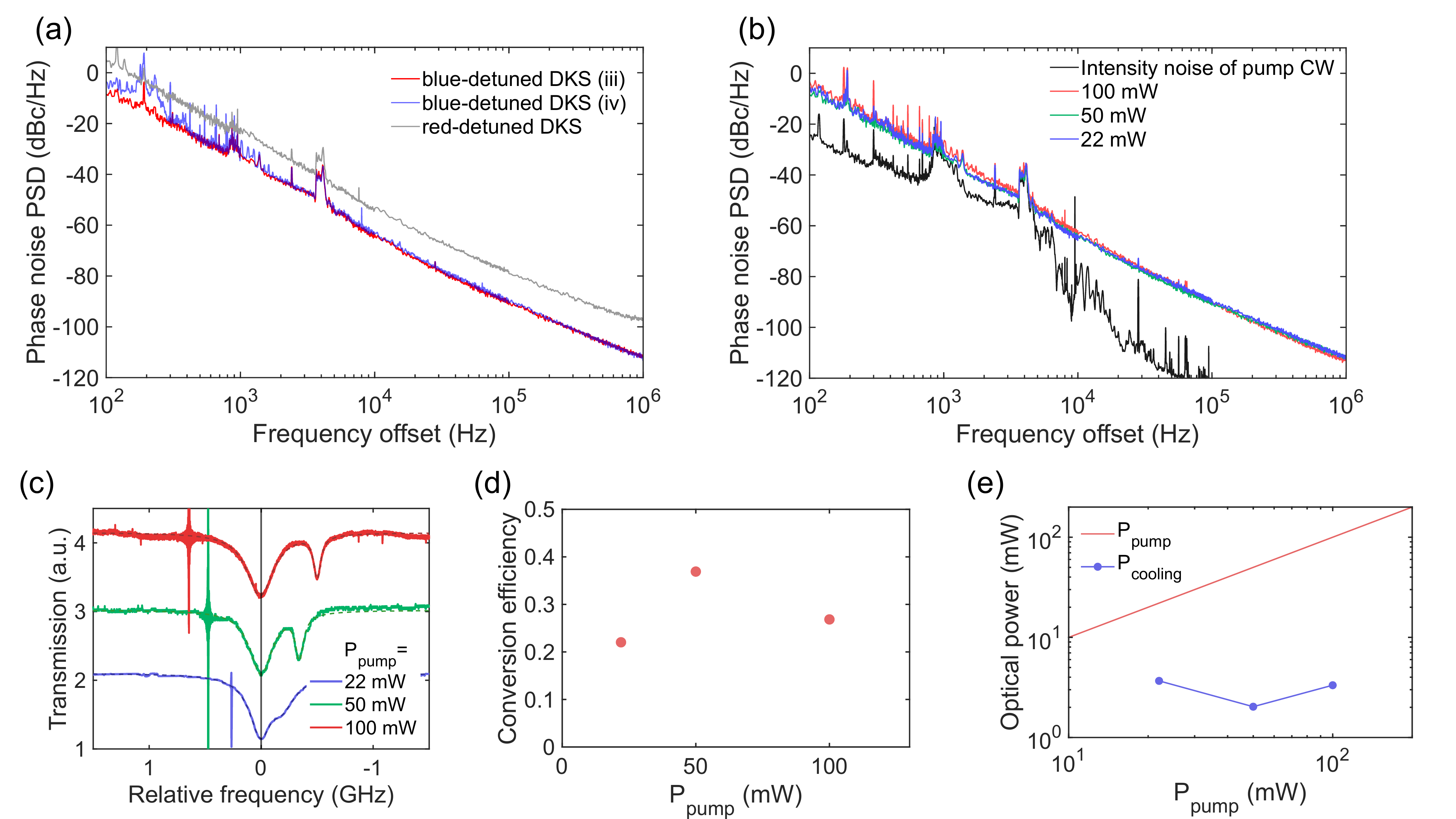}
\caption{(a) Single-sideband phase-noise power spectral density (PSD) of $f_{\rm rep}$ for blue-detuned DKSs (red and blue). Labels iii and iv correspond to those in Fig.~4. The phase noise of $f_{\rm rep}$ for a red-detuned DKS is also shown (gray). (b) Phase noise of $f_{\rm rep}$ for blue-detuned DKSs at different $P_{\rm pump}$. The phase-noise floor set by the intensity noise of the pump CW laser is also shown (black). (c) Transmission spectra around the pumped resonance in the self-cooling state for each $P_{\rm pump}$. The sharp spectral lines are beat signals between the probe CW laser and the pump CW laser. (d) Pump-to-comb power-conversion efficiency in the self-cooling state for each $P_{\rm pump}$. (e) Effective cooling power $P_{\rm cooling}$ (blue) as a function of pump power, calculated from the data in (c) and (d). The red line represents $P_{\rm pump}$ with unit slope for comparison.}
\end{figure*}

\section{Phase noise reduction via self-cooling}
Figure~5(a) shows the single-sideband phase-noise power spectral density (PSD) of $f_{\rm rep}$ measured using a two-wavelength delayed self-heterodyne interferometer (TWDI) \cite{kuse2019control, kwon2022ultrastable}. The phase noise of the blue-detuned DKSs (iii and iv, highlighted in green in Fig.~4(a)) is reduced by up to $14.5~\mathrm{dB}$ compared with that of a conventional red-detuned DKS (gray curve in Fig.~5(a)). The red-detuned DKS is generated from the same coupled-ring microresonator with $\varepsilon = 0$ via fast frequency scanning \cite{stone2018thermal, kuse2019control}. This phase-noise reduction is a distinctive feature of the blue-detuned DKS, and self-cooling works effectively as explained in Section~II because $\alpha_{\rm pump} < 0$ and $\alpha_{\rm comb} > 0$ are satisfied. Figure~5(b) shows the phase-noise PSD of $f_{\rm rep}$ for blue-detuned DKSs at various values of $P_{\rm pump}$. The details of blue-detuned DKS generation and state transitions at values other than $P_{\rm pump} = 50~\mathrm{mW}$ (used in Fig.~4) are reported in Supplementary Section~2. As shown in Fig.~5(b), the phase noise is almost identical and independent of $P_{\rm pump}$ over the range from $22$ to $100~\mathrm{mW}$. Note that $\varepsilon$ is properly adjusted to access the blue-detuned DKS when $P_{\rm pump}$ is varied. The lower limit of the phase noise set by the intensity noise of the pump CW laser is also plotted and is lower than the phase noise of the blue-detuned DKSs. Therefore, the results indicate that the self-cooling effect is saturated, although a clear correlation between cooling-laser power and phase-noise suppression has been reported in previous studies \cite{drake2020thermal}. To investigate the origin of this saturation, we evaluated the effective optical power from the pump CW laser that contributes to cooling. Figure~5(c) shows the transmission spectra around the pumped resonance, the corresponding fitting curves composed of two Lorentzian functions, and the beat signals with the pump CW laser under the self-cooling state at each $P_{\rm pump}$. These measurements were performed using a probe laser injected in the direction opposite to the pump CW laser (see Fig.~2). When the pump power is increased, the frequency separation between the pump CW laser and the pumped resonance increases, resulting in coupling ratios $C$ of $18.9\%$, $6.2\%$, and $4.6\%$ for $P_{\rm pump} = 22$, $50$, and $100~\mathrm{mW}$, respectively. The coupling ratios are calculated from the intersection between each Lorentzian fit and the center of the beat signal between the pump and probe CW lasers. The Lorentzian-shaped spectral component that appears on the red side of the pumped resonance is attributed to the soliton resonance \cite{guo2017universal} generated by the DKS pulses. The decrease in $C$ with increasing $P_{\rm pump}$ is consistent with the balance between parametric gain and cavity loss in the DKS comb. Figure~5(d) shows the measured pump-to-comb power-conversion efficiency $\eta_{\rm comb}$ for each $P_{\rm pump}$ under the self-cooling state. The optical spectra of the blue-detuned DKS measured at each $P_{\rm pump}$ are provided in Supplementary Section~2. The effective cooling power $P_{\rm cooling}$ is calculated using $P_{\rm cooling} = C P_{\rm pump} \left(1 - \eta_{\rm comb}\right)$, as shown in Fig.~5(e), under the assumption that the intracavity comb power equals the out-coupled comb power. In Fig.~5(e), $P_{\rm pump}$ increases from $22$ to $100~\mathrm{mW}$, as indicated by the red reference line with unit slope for comparison. By contrast, $P_{\rm cooling}$ plateaus over the varied $P_{\rm pump}$, which indicates saturation of the cooling effect for noise suppression.


\section{Long-term stability of self-cooling blue-detuned DKS}
Finally, we demonstrate the long-term stabilization of $f_{\rm rep}$ in the self-cooling state. In this experiment, we monitored the relative fluctuation of $f_{\rm rep}$ by heterodyning the DKS output with a fiber-based optical frequency comb. The blue-detuned DKS was generated using $P_{\rm pump} = 50~\mathrm{mW}$. Details of the measurement method are provided in Supplementary Section~3. Figure~6 compares the fluctuation of $f_{\rm rep}$ measured over a 30-minute for a normal red-detuned DKS (red) and the blue-detuned DKS (blue). Because it is difficult to maintain the red-detuned DKS in the same coupled-ring resonator with $\varepsilon = 0$, a single-ring resonator with the same cross section is used. The results show that for the red-detuned DKS (red), $f_{\rm rep}$ fluctuated over a range of up to $10.5~\mathrm{MHz}$ during the 30-minutes measurement. In contrast, the fluctuation range for the blue-detuned DKS (blue) was limited to $1.6~\mathrm{MHz}$. The standard deviation was also reduced from $2.2~\mathrm{MHz}$ to $0.41~\mathrm{MHz}$. Therefore, self-cooling is effective on both short time scales (phase noise) and long time scales (frequency fluctuation).

\begin{figure}[h]
\centering\includegraphics[width=1.0\linewidth]{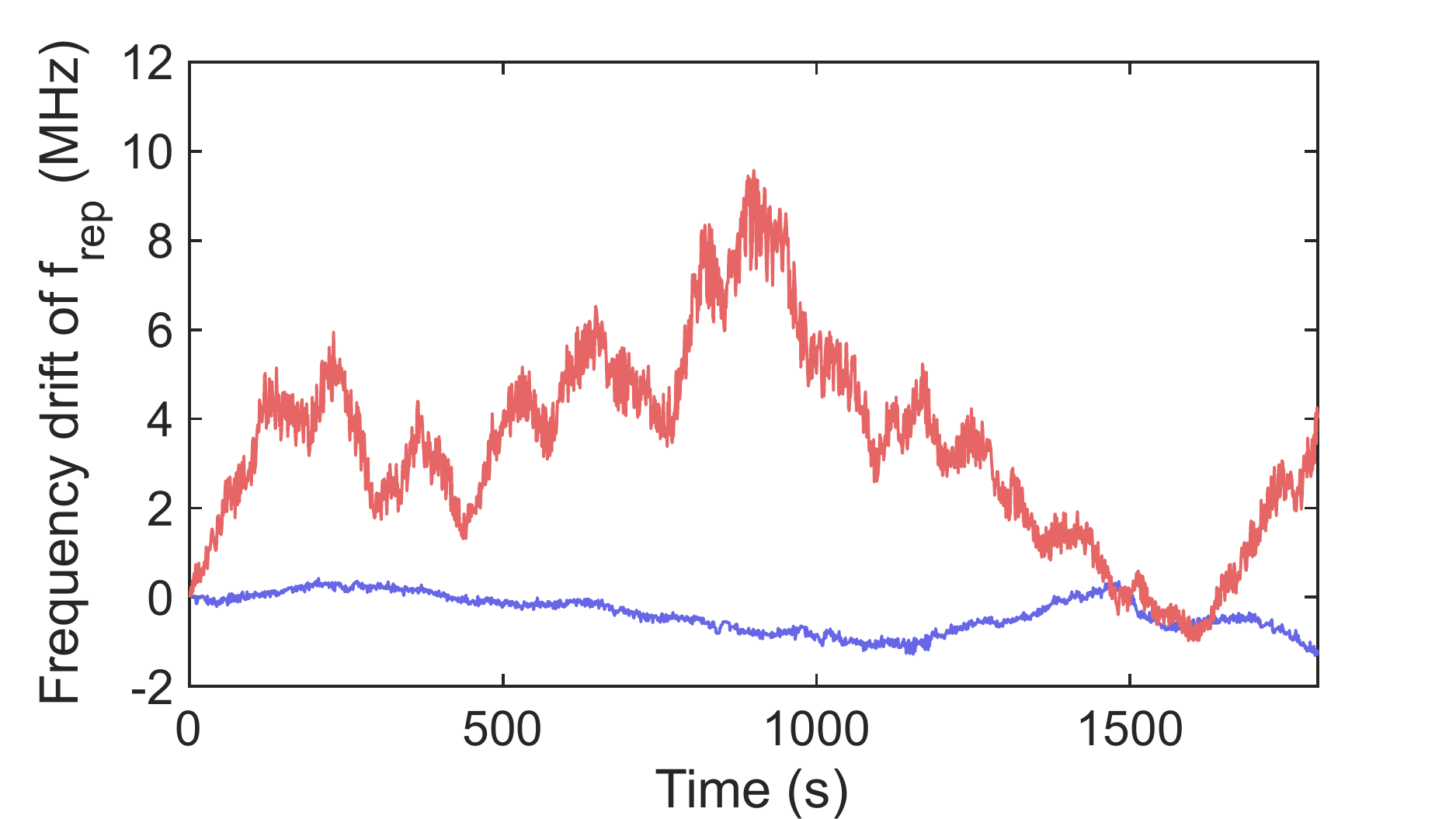}
\caption{Frequency fluctuation of $f_{\rm rep}$ measured over 30 minutes. The results for a red-detuned DKS without AMX (red) and a self-cooling blue-detuned DKS (blue) are shown.}
\end{figure}

\section{Discussion and conclusions}
We compare our method with other approaches that reduce phase noise using a blue-detuned cooling laser. Unlike methods that employ a second optical carrier on the blue side of the pumped resonance \cite{drake2020thermal, nishimoto2022thermal}, our method avoids the use of an additional optical carrier. This not only simplifies the overall configuration but also eliminates noise transfer and unwanted comb generation from a second laser. Similar to our method, Ref.~\cite{lei2022thermal} uses the pump CW laser itself as an effective cooling laser; in that scheme, different mode families are precisely adjusted so that one supports DKS generation and the other provides cooling. However, precisely aligning the resonances of different mode families at a specific frequency is challenging due to fabrication errors, even if the waveguide cross section is well designed. By contrast, our method reliably enables phase-noise reduction at the selected pump wavelength. Furthermore, because a blue-detuned DKS is used, the conversion efficiency from the pump CW laser to the DKS is much higher than that achieved with conventional red-detuned DKSs reported previously.

Although we only present the phase-noise reduction in $f_{\rm rep}$, the phase noise of individual comb modes is also expected to be suppressed. This is because the phase noise of the comb modes is limited by that of $f_{\rm rep}$ when a low-noise pump CW laser (for example, an external-cavity diode laser or a fiber laser with a linewidth of a few kilohertz) is used \cite{nishimoto2020investigation, lei2022optical}. Moreover, for applications that require scanning of comb modes such as LiDAR \cite{riemensberger2020massively}, a blue-detuned DKS has the advantage that its optical spectrum does not change significantly during frequency scanning. This occurs because the detuning varies less than the pump-laser frequency, since the detuning is thermally dragged by the pump CW laser.

Our method is based on a localized dispersion shift of the pumped resonance introduced through AMX, which enables the generation of a blue-detuned DKS. To achieve this, we employ a coupled-ring microresonator. However, similar self-cooling states are expected to be achievable in microresonators with photonic crystal structures \cite{yu2021spontaneous}.

In summary, we demonstrate the generation of a blue-detuned DKS with reduced phase noise from a coupled-ring microresonator. By adjusting the coupling strength between resonances of two rings at the pump CW laser through AMX, a blue-detuned DKS is deterministically generated. Because the pump CW laser is located on the blue side of the pumped resonance, the phase noise of $f_{\rm rep}$ is passively suppressed through self-cooling. In this demonstration, a phase-noise reduction exceeding $10~\mathrm{dB}$ is observed. Taking advantage of the intrinsic property of blue-detuned DKSs, the conversion efficiency from the pump CW laser to the DKS reaches approximately $37\%$. These properties indicate that leveraging the self-cooling characteristics of blue-detuned DKSs can strongly facilitate the development of systems that are structurally simple, robust, and energy efficient. We believe this approach is particularly effective for applications \cite{sun2023applications} such as mm/THz generation and LiDAR, which demand not only high fabrication yield and compactness but also low power consumption, high signal purity, and operational stability.

\section*{Supplementary Material}
The supplementary material includes experimental results on the reproducibility and generation of blue-detuned DKS under different pump powers, as well as the experimental setup used to measure fluctuations of $f_{\rm rep}$.

\begin{acknowledgments}
This work was supported in part by the Japan Society for the Promotion of Science (23H04806, 24K01386, 25H01887, J-PEAKS); Cabinet Office, Government of Japan; Murata
Science and Education Foundation; SCAT Foundation; Iketani Science and Technology Foundation. K. N thanks the financial support by the Japan Society for the Promotion of Science.
\end{acknowledgments}

\bibliography{Blue_soliton_references}

\end{document}